# Robust chaos in autonomous time-delay system

D.S. Arzhanukhina [1] and S.P. Kuznetsov [2,3]

[1] Saratov State University, Astrakhanskaya str., 83, Saratov, 410012, Russian Federation

[2] Kotel'nikov's Institute of Radio-Engineering and Electronics of RAS, Saratov Branch, Zelenaya str., 38, Saratov, 410019, Russian Federation

[3] Department of Physics and Astronomy, University of Potsdam, Karl-Liebknecht-Street 24/25, D-14476, Potsdam-Golm, Germany

We consider an autonomous system constructed as modification of the logistic differential equation with delay that generates successive trains of oscillations with phases evolving according to chaotic maps. The system contains two feedback loops characterized by two generally distinct retarding time parameters. In the case of their equality, chaotic dynamics is associated with the Smale-Williams attractor that corresponds to the double-expanding circle map for the phases of the carrier of the oscillatory trains. Alternatively, at appropriately chosen two different delays attractor is close to torus with Anosov dynamics on it as the phases are governed by the Fibonacci map. In both cases the attractors manifest robustness (absence of regularity windows under variation of parameters) and presumably relate to the class of structurally stable hyperbolic attractors.

## *1. Introduction*

The concept of uniformly hyperbolic chaotic dynamics was advanced in mathematical theory of dynamical systems half a century ago [1-6]. It deals with invariant sets in state space of systems composed exclusively of orbits of saddle type with such a restriction that their stable and unstable manifolds do not touch each other (intersect transversally). The hyperbolic invariant sets are structurally stable. It implies their robustness in respect to (at least small) variation of functions and parameters in the dynamical equations. Henceforth, the hyperbolic dynamics may be regarded as preferable for any possible practical application of chaos and as deserving prior research in a frame of physical and technical disciplines [7,8].

One important class of the uniform hyperbolicity is Anosov dynamics in systems where the hyperbolic invariant set occupies the whole accessible phase space; usually it is considered in the context of the phase volume preserving maps or flows (conservative dynamics). Alternatively, in the context of dissipative systems, the uniformly hyperbolic attractors were introduced; their mathematical examples are Smale – Williams solenoid, Plykin type attractors, DA-attractor of

Smale.[1] Although originally they relate to artificial discrete-time systems, analogous attractors can occur as well in Poincaré maps associated with continuous time systems governed by differential equations. In the last case it is used to speak about *suspension* of these attractors provided by the appropriate flows.

Till a short time ago, realistic examples of uniformly hyperbolic dynamics which could relate to real-world systems were poorly presented and discussed in the literature. One example is a hinge mechanism with Anosov dynamics described and studied in [9]. An artificial example of suspension of Plykin-type attractor was constructed in PhD thesis of Hunt [10], but it is surely too complicated to allow a real physical implementation. Possible occurrences of suspension of Plykin-type attractors in neuron model [11] and in modified Lorenz model [12] were discussed, but no convincing data on a level of concrete equations and numerical simulation were provided.

Recently a number of implementable systems with uniformly hyperbolic attractors were advanced and studied due to efforts of Saratov group of nonlinear dynamics [13-20]. It is essential that these examples are constructed on a base of the physical rather than the mathematical toolbox exploiting such entities as oscillators, nonlinear elements, interactions, self-oscillatory and parametric excitation etc.

One particular productive method for design of systems with uniformly hyperbolic attractors is based on using time-delay nonlinear feedback loops [21-24]. Most of the examples suggested were non-autonomous systems functioning in presence of the external periodic driving [21-23]. An exception is an autonomous system with attractor of Smale-Williams type considered in Ref. [24]. The present article is inspired by that work. We construct and study numerically an autonomous time-delay system manifesting different types of uniformly hyperbolic attractors depending on two parameters of time delay.

---

[1] The abbreviation DA stands for "derived from Anosov".

## 2. Basic equations

Let us start with *logistic delay equation* offered in due time in population biology [25]:

$$\dot{r} = \mu[1 - r(t - \tau_1)]r(t). \quad (1)$$

Here the population is characterized by the positive variable $r$ evolving in time and normalized in such way that the saturation occurs at $r = 1$. A positive parameter $\mu$ is the birth rate, $\tau_1$ is the delay time characterizing lag of the effect of saturation. According to Ref. [25], under condition $\tau_1 < \pi/2\mu$ the system has a stable stationary state $r = 1$ and for $\tau > \pi/2\mu$ self-oscillations occur. At large values $\tau_1$ the generated waveform looks like a periodic sequence of pulses (Fig.1a). The period grows with the delay $\tau_1$ as $P \cong (1 + e^{\mu\tau_1})/\mu$. Estimate for the minimal level of the population in between the pulses is $r_{\min} \cong \mu\tau\exp(-e^{\mu\tau_1} + 2\mu\tau_1 - 1)$, i.e. it manifests the double exponentially decrease with increase of $\tau_1$.

Now, let us regard the variable $r$ as *squared amplitude of some oscillatory process* with frequency $\omega_0$. For this, we set $r = x^2 + y^2$ and require the new variables to satisfy the equations

$$\begin{aligned}\dot{x} &= -\omega_0 y + \tfrac{1}{2}\mu(1 - x^2(t - \tau_1) - y^2(t - \tau_1))x, \\ \dot{y} &= \omega_0 x + \tfrac{1}{2}\mu(1 - x^2(t - \tau_1) - y^2(t - \tau_1))y.\end{aligned} \quad (2)$$

Solutions of the equations (2) instead of the solitary pulses manifest successive oscillatory trains for $x$ and $y$ while the envelope evolves exactly according to (1) (Fig.1b).

Next, let us add terms proportional to $x(t-\tau_1)x(t-\tau_2) - y(t-\tau_1)y(t-\tau_2)$ and $x(t-\tau_1)y(t-\tau_2) + x(t-\tau_2)y(t-\tau_1)$ with small coefficient $\varepsilon$ in the first and the second equation

$$\begin{aligned}\dot{x} &= -\omega_0 y + \tfrac{1}{2}\mu(1 - x^2(t - \tau_1) - y^2(t - \tau_1))x + \varepsilon[x(t-\tau_1)x(t-\tau_2) - y(t-\tau_1)y(t-\tau_2)], \\ \dot{y} &= \omega_0 x + \tfrac{1}{2}\mu(1 - x^2(t - \tau_1) - y^2(t - \tau_1))y + \varepsilon[x(t-\tau_1)y(t-\tau_2) + x(t-\tau_2)y(t-\tau_1)],\end{aligned} \quad (3)$$

where we always suppose that the delays satisfy the inequality $\tau_2 \geq \tau_1$. Now, in the case of generation of pulses with low enough level of minimal amplitude between them, just these additional terms will initiate formation of each next pulse of oscillations. Due to this, the phase of the oscillations for the new-born pulse will be determined by the phases of the previous pulses via some mapping as discussed in detail in the following sections. In presence of this initiation of the oscillations the characteristic repetition period of the pulses is reduced comparing to the original system (2).

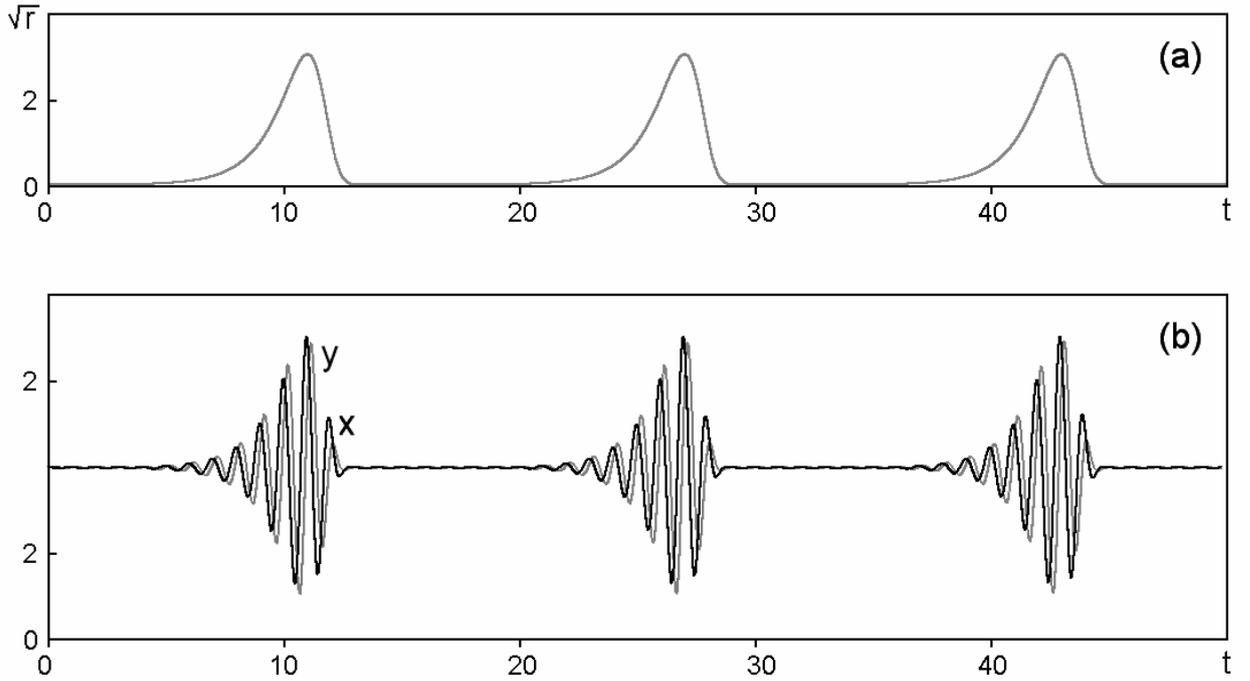

**Figure 1:** Self-pulsations in the logistic-delay equation (1) at $\mu=1.6$, $\tau_1=2$ (a) and oscillatory trains generated by system (2) at $\omega_0=2\pi$ (b). The period of the pulses is roughly $T\approx16$.

Formally, the time-delay system is infinite-dimensional. Indeed, an instantaneous state is determined by *functions* $x(t)$ and $y(t)$ defined on a time interval of length $\tau_2$. To start solution of the equations (3) the initial conditions have to be chosen as some functions $x(t)$ and $y(t)$ on the time interval $[-\tau_2, 0]$. In the course of numerical integration they are represented by arrays sampled with time step used in the integration scheme, and the elements of the arrays are replaced step by step by the newly obtained values to be accounted on latter stages of the computations. In other respects the integration schemes are analogous to those used for solution of ordinary differential equations. With arbitrary initial conditions the computations are executed for sufficiently long time and then, after reliable arrival at the attractor, processing of the data is performed including plotting waveforms, portraits of attractors, evaluation of the Lyapunov exponents etc.

Now let us turn to some concrete cases of the model (3).

### *3. Attractor of Smale – Williams type*

Let us start with the simplest particular case setting both delay time parameters equal: $\tau_2 = \tau_1 = \tau$. It leads to the equations

$$\dot{x} = -\omega_0 y + \tfrac{1}{2}\mu(1 - x^2(t-\tau) - y^2(t-\tau))x + \varepsilon[x^2(t-\tau) - y^2(t-\tau)], \quad (4)$$
$$\dot{y} = \omega_0 x + \tfrac{1}{2}\mu(1 - x^2(t-\tau) - y^2(t-\tau))y + 2\varepsilon x(t-\tau)y(t-\tau).$$

To clarify the principle of functioning of the system suppose that during some current stage of activity the variables behave as $x \sim \cos(\omega_0 t + \varphi)$ and $y \sim \sin(\omega_0 t + \varphi)$; then the additional terms are expressed as

$$x^2(t-\tau) - y^2(t-\tau) \sim \cos^2(\omega_0 t + \varphi) - \sin^2(\omega_0 t + \varphi) = \cos(2\omega_0 t + 2\varphi),$$
$$2x(t-\tau)y(t-\tau) \sim 2\cos(\omega_0 t + \varphi)\sin(\omega_0 t + \varphi) = \sin(2\omega_0 t + 2\varphi). \quad (5)$$

These relations determine actually the stimulating signal for oscillations on the immediate next stage of activity. The phase shift of this signal is transferred to the oscillations arising on this stage. Hence, the phase undergoes the doubling transformation

$$\varphi_{n+1} = 2\varphi_n + \text{const}. \quad (6)$$

It is the expanding circle map, or the Bernoulli map, which is chaotic and is characterized by the Lyapunov exponent $\Lambda = \ln 2 \approx 0.693$.

We may consider a Poincaré map for the time-delay system that corresponds to transformation of the infinite-dimensional space vector on a time interval from one pulse of oscillations to the next one, Attractor of this Poincaré map is supposed to be a kind of Smale-Williams solenoid embedded in the infinite-dimensional state space. Actually, the phase φ plays a role of the angular variable in the Smale-Williams solenoid while in other directions compression of the phase volume will take place. Respectively, attractor of the original autonomous time-delay system

(4) is a suspension of the Smale-Williams solenoid in the infinite-dimensional state space.

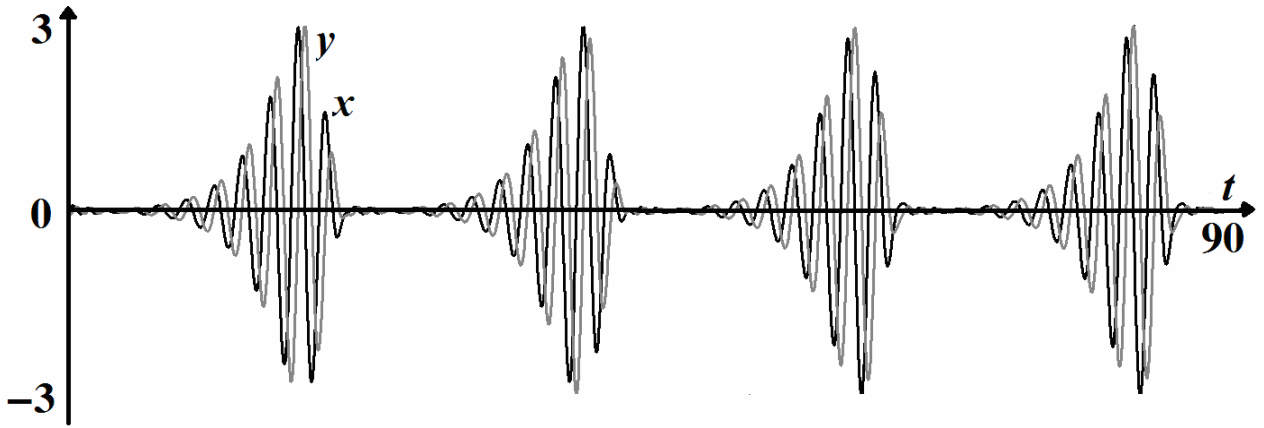

**Figure 2:** Waveforms for the dynamical variables *y* (black) and *x* (gray) according to the results of the numerical solution of equation (4) at $\omega_0=2\pi$, $\mu=1.6$, $\omega_0 = 2\pi$, $\tau= 2$, $\varepsilon = 0.05$.

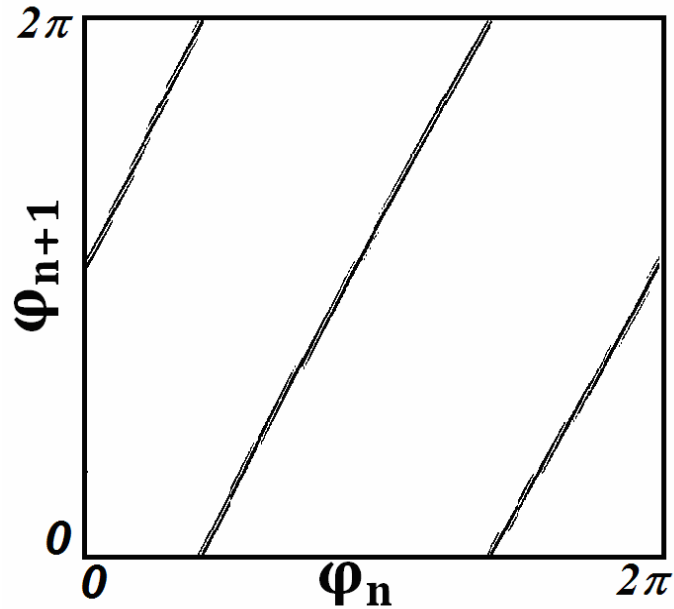

**Figure 3:** Diagram illustrating the transformation of phases in the successive stages of the activity plotted according to the results of the numerical solution of equation (4) at $\omega_0=2\pi$, $\mu=1.6$, $\omega_0 = 2\pi$, $\tau= 2$, $\varepsilon = 0.05$

Figure 2 shows waveforms for dynamical variables *x* and *y* illustrating operation of the system (4) at

$$\omega_0=2\pi,\ \mu=1.6,\ \omega_0 = 2\pi,\ \tau= 2,\ \varepsilon = 0.05. \qquad (7)$$

In accordance with the above qualitative considerations, the process looks like a sequence of trains of oscillations. Note that the average period of their appearance is less than that in models (1), (2), and in the regime under discussion it is $T\approx10$.

Although represented by nearly periodic alternation of excitation and suppression of the oscillations, the process is actually chaotic (in contrast to the regular one observed in Fig.1b). Chaos reveals itself in variations of phases on successive stages of activity. More accurate analysis shows that these phases obey approximatelt the Bernoulli map. Figure 3 illustrates it by a diagram obtained in the numerical simulation. The phases were determined at moments of maximal amplitude at successive excitation stages as $\varphi_n = \arg(x - iy)$ and plotted for sufficiently large number of processed successive oscillatory trains. Surely, the mapping for the phases looks topologically equivalent to the expected Bernoulli map: one complete bypass for the pre-image $\varphi_n$ (that is variation by $2\pi$) corresponds to the two-fold bypass for the image $\varphi_{n+1}$.

Figure 4 shows 3D and 2D projections of the attractor from the infinite-dimensional phase space of our time-delay system.

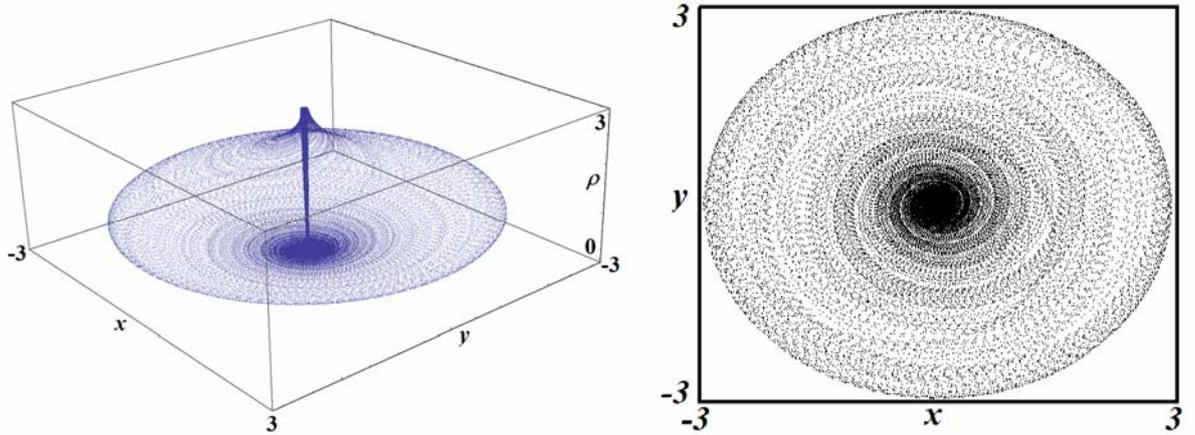

**Figure 4:** 3D and 2D projections of the attractor from the infinite-dimensional phase space at $\omega_0 = 2\pi$, $\mu = 1.6, \omega_0 = 2\pi$, $\tau = 2$, $\varepsilon = 0.05$. The first diagram is plotted in the coordinates $(x, y, \rho)$, where $\rho = \sqrt{x^2(t-\tau) + y^2(t-\tau)}$, and the second on the plane of variables $(x, y)$.

For computation of Lyapunov exponents we use the Benettin algorithm [26, 27] adapted for the time-delayed systems [28, 29]. It is based on simultaneous numerical solutions of the equations (4) and the variation equations

$$\begin{aligned}
\dot{\tilde{x}} &= -\omega_0 \tilde{y} + \tfrac{1}{2}\mu[1 - x^2(t-\tau) - y^2(t-\tau)]\tilde{x} - \mu x[x(t-\tau)\tilde{x}(t-\tau) + y(t-\tau)\tilde{y}(t-\tau)] \\
&\quad + 2\varepsilon[x(t-\tau)\tilde{x}(t-\tau) - y(t-\tau)\tilde{y}(t-\tau)], \\
\dot{\tilde{y}} &= \omega_0 \tilde{x} + \tfrac{1}{2}\mu[1 - x^2(t-\tau) - y^2(t-\tau)]\tilde{y} - \mu y[x(t-\tau)\tilde{x}(t-\tau) + y(t-\tau)\tilde{y}(t-\tau)] \\
&\quad + \varepsilon[y(t-\tau)\tilde{x}(t-\tau) + x(t-\tau)\tilde{y}(t-\tau)].
\end{aligned} \qquad (8)$$

Formally, there are an infinite number of Lyapunov exponents in the time-delayed system, but we necessarily restrict ourselves dealing with a finite number of them. To evaluate the larger $M$ exponents we integrate a collection of $M$ sets of equations of form (8). In the computations, the perturbation vectors are represented as finite-dimensional arrays of values $\tilde{x}$ and $\tilde{y}$ sampled on intervals of length $\tau$ with time step of the integration scheme of the delay equations. The dot product for the vectors involved in normalization and Gram-Schmidt orthogonalization required in the Benettin algorithm is determined as a sum of products of pairs of the elements relating to one and another array.

It is convenient to use normalization of the Lyapunov exponents $\lambda_i$ by the average period of pulse repetition $T$, namely, to set $\Lambda = \lambda T$. It makes natural to relate them with the Lyapunov exponent of the Bernoulli map obtained in the qualitative analysis.

According to the computations, in the regime corresponding to parameters (7) the largest four Lyapunov exponents are

$$\Lambda_1 = 0.6964, \ \Lambda_2 = 0.0000, \ \Lambda_3 = -12.129, \ \Lambda_4 = -15.538,... \qquad (9)$$

The estimate of the Kaplane – Yorke dimension for the attractor of the Poincare map (the Smale-Williams solenoid) is $D_{KY}^0 = 1 + \Lambda_1 / |\Lambda_3| \approx 1.06$. Dimension of the attractor in the infinite-dimensional phase space of the autonomous system is $D_{KY} = 2 + (\Lambda_1 + \Lambda_2) / |\Lambda_3| = 1 + D_{KY}^0 \approx 2.06$.

Figure 5 shows a plot for four Lyapunov exponents of the system (4) verses $\varepsilon$, $\tau$ and $\mu$. Observe that the largest exponent remains almost constant in a wide interval of the parameter variation and remains close to the expected value ln2=0.693…The second is close to zero (up to numerical inaccuracy). As usual in autonomous systems, it is interpreted as associated with perturbations of infinitesimal shift along the phase trajectory. Other exponents are negative.

Note smooth character of dependences of the Lyapunov exponents on parameters, particularly, there are no drops in the plot of $\Lambda_1$, which could correspond to regularity windows intrinsic usually to non-hyperbolic systems. It confirms robust nature of chaos, which is natural in a frame of our qualitative argumentation that it is associated with the uniformly hyperbolic attractor of Smale-Williams type.

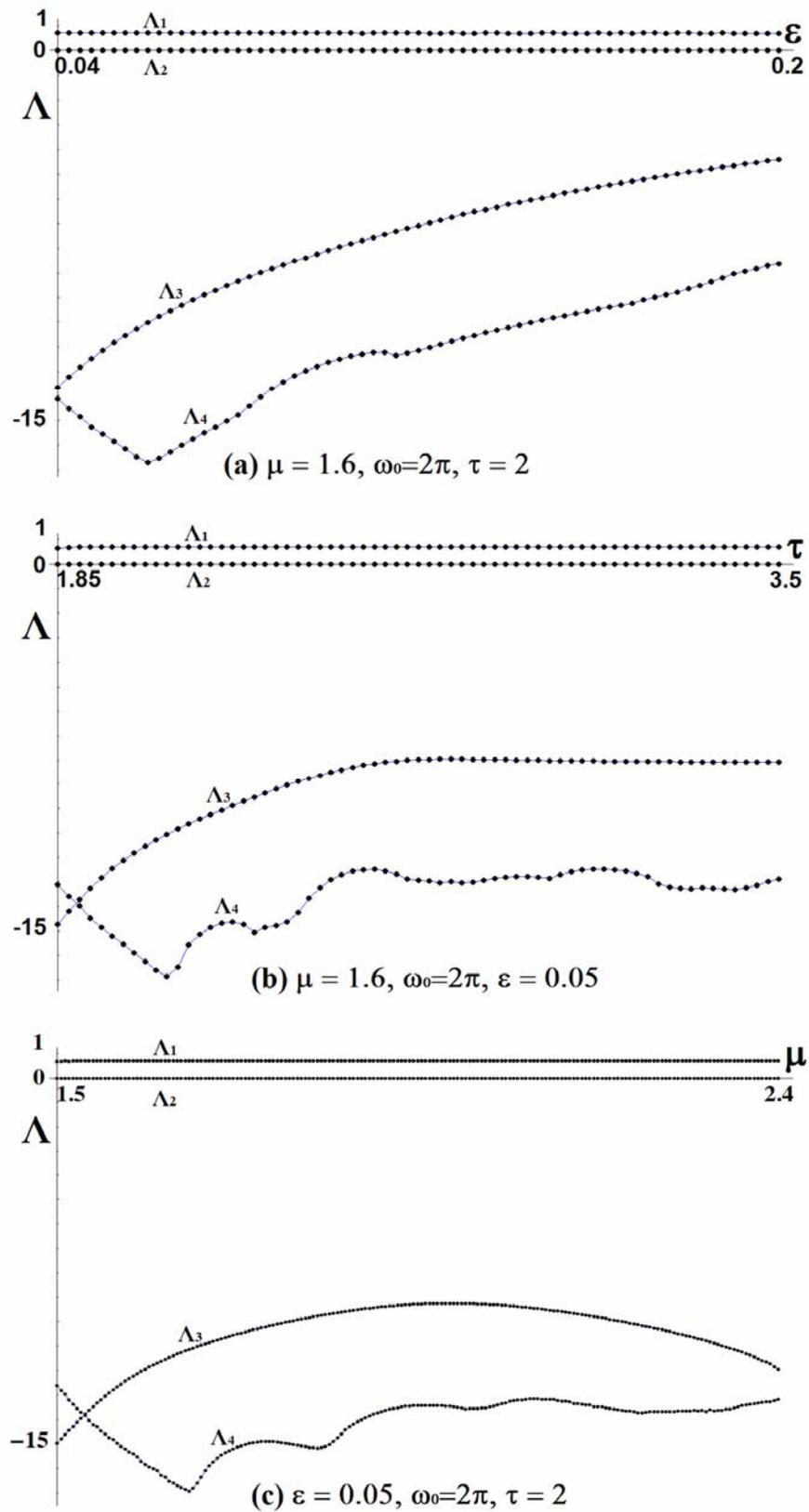

**Figure 6:** Lyapunov exponents of the system (4) versus the parameters $\varepsilon$ (a), $\tau$ (b), $\mu$ (c).

### *3. Attractor associated with with Anosov-type dynamics on torus*

Now return to the model with two different time delays $\tau_1$ and $\tau_2$:

$$\begin{aligned}\dot{x} &= -\omega_0 y + \tfrac{1}{2}\mu(1-x^2(t-\tau_1)-y^2(t-\tau_1))x + \varepsilon[x(t-\tau_1)x(t-\tau_2)-y(t-\tau_1)y(t-\tau_2)], \\ \dot{y} &= \omega_0 x + \tfrac{1}{2}\mu(1-x^2(t-\tau_1)-y^2(t-\tau_1))y + \varepsilon[x(t-\tau_1)y(t-\tau_2)+x(t-\tau_2)y(t-\tau_1)].\end{aligned} \quad (10)$$

and select these parameters to get such situation that the delayed signals stimulating excitation of a new $n+1$-th pulse of oscillations arrive being emitted from the previous two pulses $n$ (delay $\tau_1$) and $n-1$ (delay $\tau_2$).

Suppose that at the previous two activity stages the phases were $\varphi_n$ and $\varphi_{n-1}$, namely, $x(t-\tau_1) \sim \cos(\omega_0 t + \varphi_n)$, $y(t-\tau_1) \sim \sin(\omega_0 t + \varphi_n)$, $x(t-\tau_2) \sim \cos(\omega_0 t + \varphi_{n-1})$, $yx(t-\tau_2) \sim \sin(\omega_0 t + \varphi_{n-1})$. Then, the additional terms in the right-hand parts are expressed as

$$\begin{aligned}x(t-\tau_1)x(t-\tau_2)-y(t-\tau_1)y(t-\tau_2) &\sim \cos(2\omega_0 t + \varphi_n + \varphi_{n-1}), \\ x(t-\tau_1)y(t-\tau_2)+x(t-\tau_2)y(t-\tau_1) &\sim \sin(2\omega_0 t + \varphi_n + \varphi_{n-1}).\end{aligned} \quad (11)$$

As follows, on the next, $(n+1)$-th stage of activity the phase will be determined (up to a constant) by the *Fibonacci map*

$$\varphi_{n+1} = \varphi_n + \varphi_{n-1} + \text{const} \ (\text{mod}\, 2\pi). \quad (12)$$

Of what kind attractor is in this case? For the infinite-dimensional Poincaré map of our time-delay system attractor is an object close geometrically to a two-dimensional torus, and the discrete-time dynamics of the angular variables on this torus obey the map (12). It relates to the class of Anosov maps. Due to structural stability of the Anosov dynamics, one can conjecture that on the attractor in the infinite-dimensional state space of the Poincaré map the dynamics remains of the same nature; in other dimensions the phase volume compression occurs that corresponds to the approach of orbits to the attractor. Respectively, attractor in the phase space of the continuous-time system (10) is a suspension of that object, and here (in the autonomous system) we have an additional neutral direction associated with a zero Lyapunov exponent.

Figure 3 shows waveforms for the variables *x* and *y* obtained from numerical solution of the equations (10) at

$$\mu = 1.6,\ \varepsilon = 0.02,\ \omega_0 = 2\pi, \tau_1 = 2,\ \tau_2 = 14. \qquad (13)$$

The process looks like a sequence of trains of oscillations for which the phases of the high-frequency carrier vary in a random-like way from one to the next stage of activity. The average period of repetition of the pulses evaluated numerically is roughly $T \approx 10.85$.

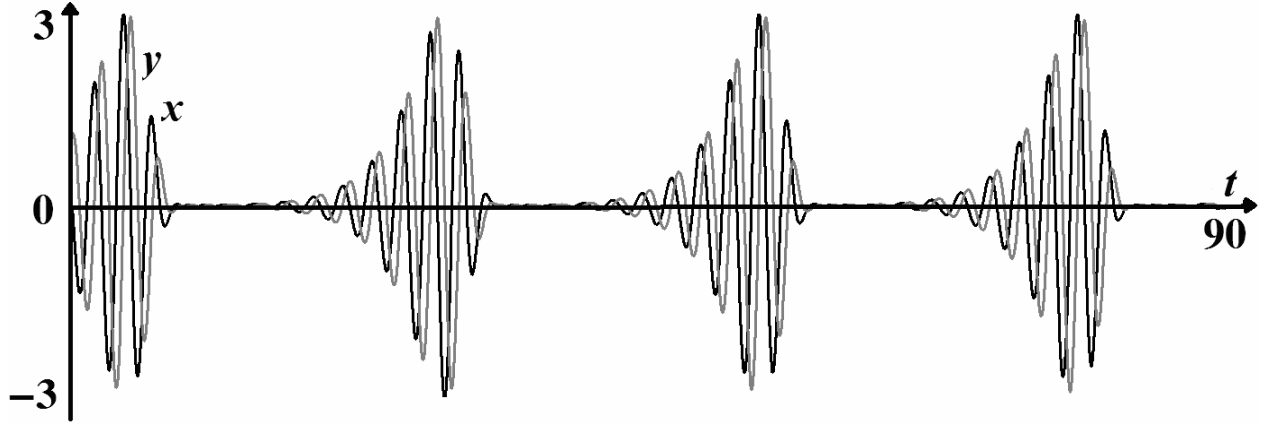

**Figure 3:** Waveforms for variables $x$ (black) and $y$ (gray) according to the numerical solution of equation (10) at $\mu = 1.6,\ \varepsilon = 0.02,\ \omega_0 = 2\pi, \tau_1 = 2,\ \tau_2 = 7\tau_1 = 14$.

Figure 4 shows 3D and 2D projections of the attractor from the infinite-dimensional phase space of our time-delay system on the plane of variables $(x, y)$.

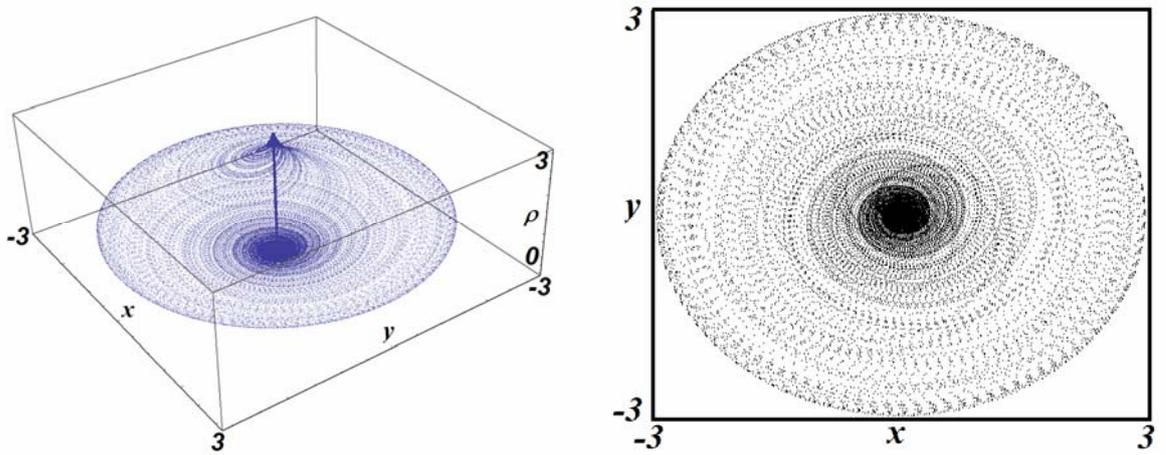

**Figure 4:** 3D and 2D projections of the attractor from the infinite-dimensional phase space at $\mu = 1.6,\ \varepsilon = 0.02,\ \omega_0 = 2\pi, \tau_1 = 2,\ \tau_2 = 7\tau_1 = 14$. The first diagram is plotted in the coordinates $(x, y, \rho)$, where $\rho = \sqrt{x^2(t-\tau) + y^2(t-\tau)}$, and the second on the plane of variables $(x, y)$.

Figure 7 illustrates correspondence of the dynamics of the phases to the Fibonacci map. The phases $\varphi_n$ are determined at instants of maximal amplitude in the oscillation trains according to relation $\varphi = \arg(x - iy)$. Diagram (a) presents the data in coordinates $(\varphi_n + \varphi_{n-1}, \varphi_{n+1})$ and one can see that the dots lie in a strip parallel to the bisector. (Although the strip looks widened, this does not violate the expected topological nature of the mapping.) Diagram (b) shows the 3D diagram for the phases in coordinates $(\varphi_{n+1}, \varphi_n, \varphi_{n-1})$, which corresponds visually to the Fibonacci map.

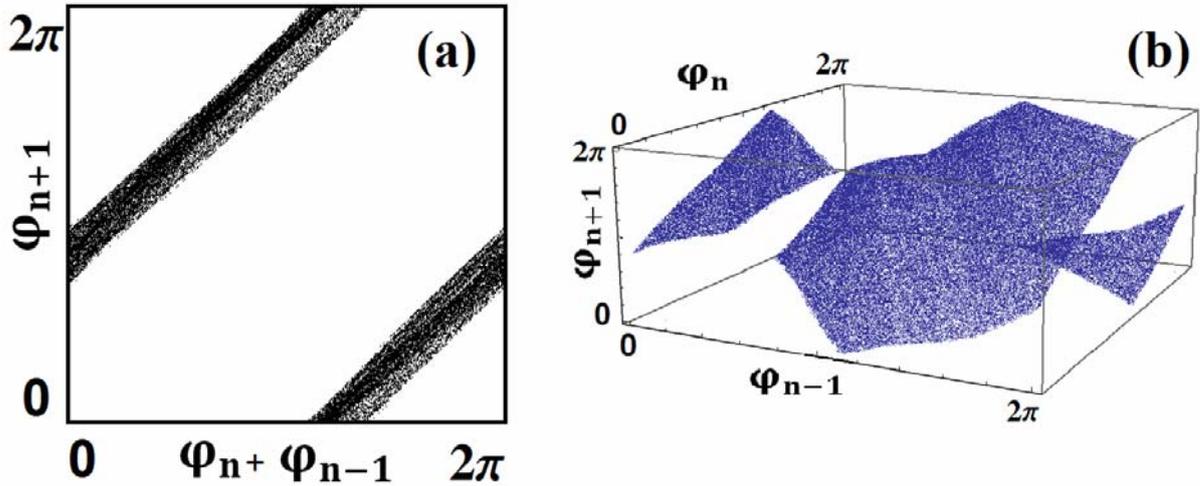

**Figure 7:** Diagrams illustrating dynamics of the phases of the system (10) in successive stages of activity.

One more way to illustrate the Anosov dynamics for the phases is similar to the famous picture of Arnold's cat [30, 15, 16]. For each pair of phases $\varphi_n$ and $\varphi_{n-1}$ on the *n*-th and *n*–1 stages of activity, we determine either the point with such coordinates fit or not in the picture of the cat face drawn on the plane. If yes, then the point is marked on the graph, and on the following two plots the corresponding points are depicted, respectively after 3 and 6 iteration steps. If the initial point does not fall within the specified cat face area, the dots are not marked, and we carry on with further iterations. Figure 8 shows the pictures obtained by this method comparing with respective diagram obtained from iterations of the Fibonacci map itself.

For computation of the Lyapunov exponents we use the algorithm similar to that described in the previous section. It is based on the joint numerical solutions of the equations (10) and the variation equations

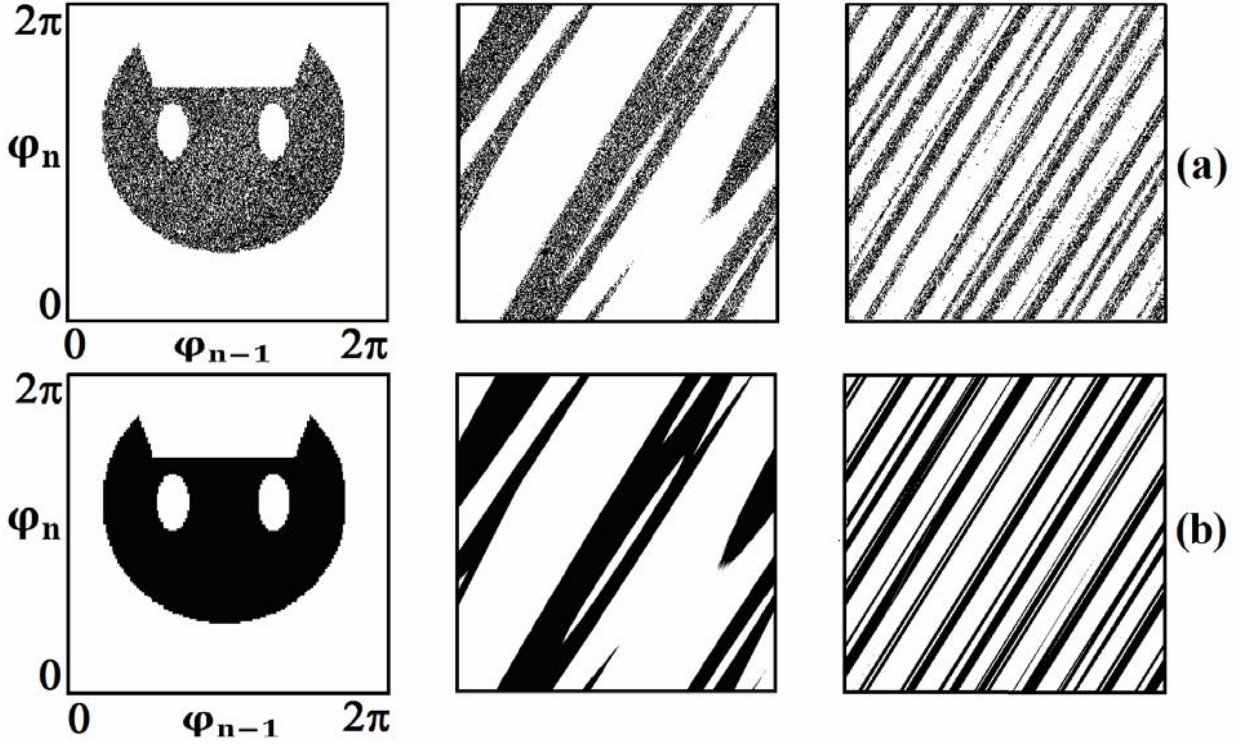

**Figure 8:** Transformation of the cat face area under successive iterations of the mapping for the phases obtained in computations for the model (16) (a) and for the Fibonacci map (b). (See explanations in the text.)

$$\dot{\tilde{x}} = -\omega_0 \tilde{y} + \tfrac{1}{2}\mu(1 - x^2(t-\tau_1) - y^2(t-\tau_1))\tilde{x} - \mu x(x(t-\tau_1)\tilde{x}(t-\tau_1) + y(t-\tau_1)\tilde{y}(t-\tau_1)) +$$
$$+ \varepsilon[x(t-\tau_1)\tilde{x}(t-\tau_2) + x(t-\tau_2)\tilde{x}(t-\tau_1) - y(t-\tau_1)\tilde{y}(t-\tau_2) - y(t-\tau_2)\tilde{y}(t-\tau_1)],$$
(14)
$$\dot{\tilde{y}} = \omega_0 \tilde{x} + \tfrac{1}{2}\mu(1 - x^2(t-\tau_1) - y^2(t-\tau_1))\tilde{y} - \mu y(x(t-\tau_1)\tilde{x}(t-\tau_1) + y(t-\tau_1)\tilde{y}(t-\tau_1))$$
$$+ \varepsilon[y(t-\tau_2)\tilde{x}(t-\tau_1) + y(t-\tau_2)\tilde{x}(t-\tau_2) + x(t-\tau_2)\tilde{y}(t-\tau_1) + x(t-\tau_1)\tilde{y}(t-\tau_2)].$$

The perturbation vectors are represented in computations as finite arrays of $\tilde{x}$ and $\tilde{y}$ on time intervals of length $\tau_2$ sampled with the integration step used for the numerical solution of the equations. We compute four larger exponents and normalize them by the factor $T$; at parameters (13) they are

$$\Lambda_1 = 0.4851, \ \Lambda_2 = 0.0003, \ \Lambda_3 = -0.4691, \ \Lambda_4 = -0.5404 \ldots \quad (15)$$

The first and the third Lyapunov exponents are close in magnitude and opposite in sign. Their may be compared with the Lyapunov exponents of the Fibonacci map $\Lambda = \pm \ln(1+\sqrt{5})/2 = \pm 0.4812\ldots$. The second exponent is zero up to numerical inaccuracy. The fourth and subsequent Lyapunov exponents are all negative, of larger absolute values.

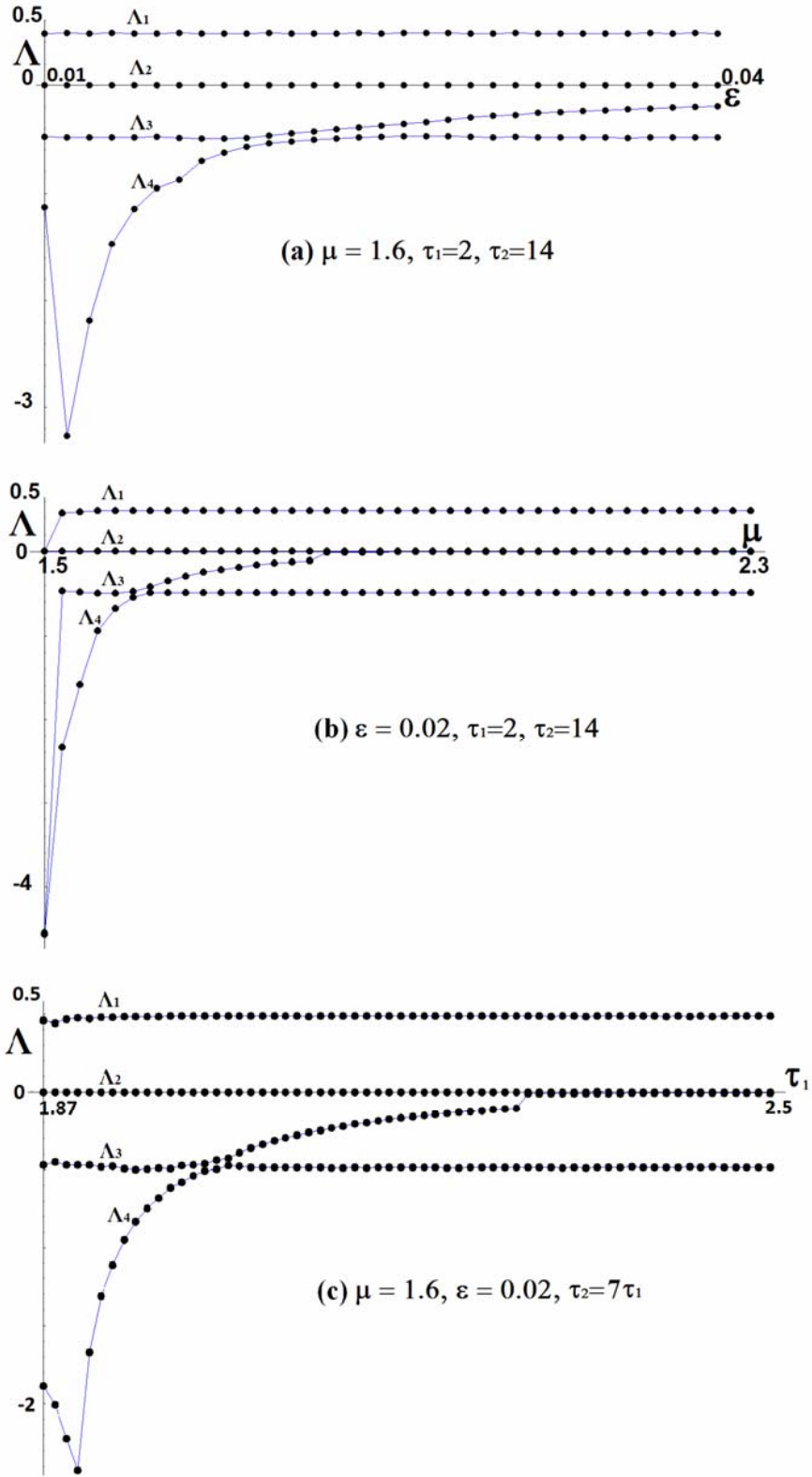

**Figure 9:** Lyapunov exponents of the system (10) depending on the parameters $\varepsilon$, $\mu$ and $\tau$.

Figure 9 shows plots for the Lyapunov exponents depending on the coupling parameter $\varepsilon$ (a), parameter of activity $\mu$ (b) and the delay $\tau_1$ (c) at fixed other parameters. As seen, in a fairly wide range of the parameters the largest Lyapunov exponent remains approximately constant and among the negative exponents there is one close in the absolute value to the positive exponent. It indicates persistence of the Anosov dynamics on the attractor.

Viewing the graphics in Fig. 9 one can distinguish two somewhat different situations. Namely, in the left parts of the diagrams the first negative exponent is nearly equal to the positive one in absolute value, and the next negative exponent is lesser. As we go to the right some crossover occurs, and the two negative exponents exchange their places. After that, obviously, approach of trajectories to the attractor in time is slower than the convergence and the divergence of orbits on the attractor itself. (Apparently, it implies different nature of the attractors in these two situations, but we leave the comparative analysis of them outside the frame of the present article.)

We outline again the smooth character of dependences of the Lyapunov exponents on parameters that confirms robust nature of chaos. It may be thought that the attractors we deal with in this section are uniformly hyperbolic, close to the so-called "non-strange chaotic attractors" briefly mentioned and discussed in the literature [31-33].

## *Conclusion*

In this paper, we introduce an autonomous system built on the basis of the logistic differential equation with delay. As shown, the dynamics of the phases of generated successive oscillatory trains corresponds to expanding circle map or Anosov map on a torus depending on selection of two time retarding parameters. Numerical simulations show robust chaos generation in these cases, and we conjecture the uniformly hyperbolic nature on the attractors in the infinite-dimensional state space of the time-delay system.

The suggested system and analogous constructions may be realized on a basis of electron devices as generators of robust chaos; physically they may be simpler in implementation comparing, for example, with systems constructed on the basis of coupled oscillators [13-16].

Due to the insensitivity of chaos to variations in parameters, the chaos generators of this kind are of interest from a practical point of view, for example in application for communication, random number generation, cryptographic schemes etc. [7].

D.S.A. acknowledges support of this work by RFBR grant № 14-02-31162, and S.P.K. thanks Prof. A. Pikovsky for discussions.